\documentclass[manuscript,11pt]{aastex}
\input blackdvi.tex

\def\au{{\rm\,AU}}

\def\mearth{{\,M_\oplus}}

\def\10^#1{\times 10^{#1}}

\def\gapprox{\hbox{ \lower0.5ex\hbox{$\sim$} \kern-1.1em 
                 \raise0.5ex\hbox{$>$} }} 
\def\lapprox{\hbox{ \lower0.5ex\hbox{$\sim$} \kern-1.1em 
                 \raise0.5ex\hbox{$<$} }} 

\lefthead{Levison \& Morbidelli}
\righthead{Kuiper Belt}

\begin{document}

\title{Models of the Collisional Damping Scenario for \Red{Ice Giant
    Planets and} Kuiper Belt Formation}

\author{Harold F. Levison}
\affil{Department of Space Studies, Southwest Research Institute,
       Boulder, CO, USA 80302}
\authoremail{hal@boulder.swri.edu}

\and

\author{Alessandro  Morbidelli}
\affil{Observatoire de la C\^ote d'Azur, Nice, France}

\newpage

\begin{abstract}
  
  Chiang et al$.$~(2006, hereafter C06) have recently proposed that
  the observed structure of the Kuiper belt could be the result of a
  dynamical instability of a system of $\sim\!5$ primordial ice giant
  planets in the outer Solar System.  According to this scenario,
  before the instability occurred, these giants were growing in a
  highly collisionally damped environment according to the arguments
  in Goldreich et al$.$~(2004a,b, hereafter G04).  Here we test this
  hypothesis with a series of numerical simulations using a new code
  designed to incorporate the dynamical effects of collisions.  We
  find that we cannot reproduce the observed Solar System.  In
  particular, G04 and C06 argue that during the instability, \Red{all
    but two} of the ice giants would be ejected from the Solar System
  by Jupiter and Saturn, leaving Uranus and Neptune behind.  We find
  that ejections are actually rare and that instead the systems spread
  outward.  This always leads to a configuration with too many planets
  that are too far from the Sun.  Thus, we conclude that both G04's
  scheme for the formation of Uranus and Neptune and C06's Kuiper belt
  formation scenario are not viable in their current forms.

\end{abstract}

\medskip
\medskip
\section{Introduction}
\label{sec_into}

The investigation of the primordial processes that sculpted the
structure of the Kuiper belt is still an active topic of research.
Several models have been developed over the last decade, based on the
effects of Neptune's migration on the distant planetesimal disk
(Malhotra~1995; Hahn \& Malhotra~1999; Gomes~2003; Levison \&
Morbidelli~2003; see Morbidelli et al$.$ 2003 for a review). However,
many aspects of the Kuiper belt have not yet been fully explained.
Moreover, a new paradigm about the giant planets orbital evolution has
recently been proposed (Tsiganis et al$.$ 2005; Gomes et al$.$ 2005;
see Morbidelli~2005 or Levison et al$.$~2006 for reviews), which calls
for a global revisiting of the Kuiper belt sculpting problem.

In this evolving situation, Chiang et al$.$~(2006, hereafter C06) have
recently proposed an novel scenario. The idea is based on a recent
pair of papers by Goldreich et al$.$~(2004a,b, hereafter G04 for the
pair and G04a and G04b for each individual paper), who, based on
analytic arguments, predicted that originally \Red{roughly five}
planets began to grow between $\sim\!20$ and $\sim\!40\,$AU.  However,
as these planets grew to masses of $\sim\!15\,M_\oplus$ their orbits
went unstable, \Red{all but two} of them were ejected, leaving Uranus
and Neptune in their current orbits.  C06 argued that this violent
process could explain the structure of the Kuiper belt that we see
today.  We review the C06 scenario in more detail in
section~\ref{sec_chiang}.  Like G04, the C06 scenario was not tested
with numerical simulations, but was solely supported by
order-of-magnitude analytic estimates, which were only possible under
a number of simplifications and assumptions.

Therefore, the goal of this paper is to simulate numerically C06's
scenario, in order to see if the presence of \Red{five} Neptune mass
bodies (or a similar configuration) in a primordial planetesimal disk
is indeed consistent with the observed structure of the outer Solar
System (the orbital distribution of the Kuiper belt and of the
planets).  Because G04 and C06 scenarios heavily rely on the presence
of a highly collisional planetesimal disk, we need first to develop a
new numerical integrator that takes collisions into account as well as
their effects on the dynamical evolution.  This code is described and
tested in Section~\ref{sec_code}. In section~4 we then describe the
results of the simulations that we did of C06's scenario. The
conclusions and the implications are discussed in section~5.

\medskip
\medskip
\section{A Brief Description of the G04/C06 Scenario}
\label{sec_chiang}

As we discussed above, C06's scenario for sculpting the Kuiper belt is
built on G04's scheme for planet formation.  In G04, the authors
pushed to an extreme the concept of runaway (Ohtsuki \& Ida 1990;
Kokubo \& Ida 1996) and oligarchic growth (Kokubo \& Ida 1998; Thommes
et al$.$ 2003; Chambers, 2006) of proto-planets in planetesimal disks.
Unlike previous works, they assumed that the bulk of the mass of the
planetesimal disk is in particles so small (sub-meter to cm in size)
that they have very short mean free paths. In this situation the disk
is highly collisional, and the collisional damping is so efficient
that the orbital excitation passed from the growing planets to the
disk is instantaneously dissipated.  With this set-up, the extremely
cold disk exerts a very effective and time enduring dynamical friction
on the growing planetary embryos, whose orbital eccentricities and
inclinations remain very small.  Consequently, the embryos grow
quickly, accreting the neighboring material due to the fact that
gravitational focusing is large.

The order-of-magnitude analytic estimates that describe this evolution
lead to the conclusion that the system reaches a steady state
consisting of a chain of planets, separated by 5 Hill radii embedded
in a sea of small particles.  As the planets grow, their masses
increase while their number decreases.  This process continues until
the surface density of the planetary embryos, $\Sigma$, is equal to
that of the disk, $\sigma$.  If the mass of the disk is tuned to
obtain planets of Uranus/Neptune mass when $\Sigma\!\sim\!\sigma$ then
the conclusion is that about 5 of these planets had to form in the
range 20--$40\,$AU.

G04 argue that when $\Sigma\!\sim\!\sigma$ the dynamical friction
exerted on the planets by the disk is no longer sufficient to
stabilize the planetary orbits.  Consequently, the planets start to
scatter one another onto highly elliptical and inclined orbits. They
assume that \Red{all but two of the original ice planets (i$.$e$.$
  three of five in the nominal case)} are ejected from the Solar
System in this scattering process (no attempts were made to model this
event).  Once \Red{their companions} have disappeared, the two
remaining planets feel a much weaker excitation, and therefore their
orbits can be damped by the dynamical friction exerted by the
remaining disk.  \Red{This damping phase is then followed by a period
  of outward migration (Fern\'andez \& Ip~1986). They therefore}
become Uranus and Neptune, with quasi-circular co-planar orbits at
$\sim\!20$ and $\sim\!30$ AU.

C06 argues that this basic scenario, with some small modifications,
can explain much of the structure currently seen in the Kuiper belt.
The Kuiper belt displays a very complex dynamical structure.  For our
purposes, four characteristics of the Kuiper belt are important: 1)
The Kuiper belt apparently ends near $50\,$AU (Trujillo \& Brown~2001;
Allen et al$.$~2001, 2002).  2) The Kuiper belt appears to consist of
at least two distinct populations with different dynamical and
physical properties (Brown~2001; Levison \& Stern~2001; Trujillo \&
Brown~2002; Tegler \& Romanishin,~2003).  One group is dynamically
quiescent and thus we call it the {\it cold population}.  All the
objects in this population are red in color.  The other group is
dynamically excited, inclinations can be as large as 40$^\circ$, and
thus we call it the {\it hot population}.  It, too, contains red
objects, but it also contains about as many objects that are gray in
color.  The largest objects in the Kuiper belt reside in the hot
population. 3) Many members of the hot population are trapped in the
mean motion resonances (MMRs) with Neptune.  The most important of
these is the 2:3MMR, which is occupied by Pluto. 4) The Kuiper belt
only contains less than roughly $0.1\,M_\oplus$ of material (Jewitt et
al.~1996; Chiang \& Brown~1999; Trujillo et al.~2001; Gladman et
al.~2001; Bernstein et al$.$~2004).  This is surprising given that
accretion models predict that $\gtrsim 10\,M_\oplus$ must have existed
in this region in order for the objects that we see to grow
(Stern~1996; Stern \& Colwell,~1997; Kenyon \& Luu,~1998, 1999).

C06 suggests the following explanation for the Kuiper belt's
structure.  First, in order to make the edge (characteristic~1 above),
they assume that the planetesimal disk is truncated at $\sim\!47\,$AU.
This assumption is legitimate given the work of Youdin \& Shu (2002)
and Youdin \& Chiang (2004) on planetesimal formation.  They assume
--as a variant of the pure G04 scenario -- that some coagulation
actually occurred in the planetesimal disk while the planets were
growing. This coagulation produced a population of objects with a size
distribution and a total number comparable to the hot population that
we see today.  Because this population constitutes only a small
fraction of the total disk's mass, their existence does not change the
overall collisional properties of the disk, which are essential for
G04's story.

Thus, during the final growth of the ice giants, C06 envisions three
distinct populations: the `planetary embryos' (objects that eventually
become Neptune-sized), the `KBOs' (macroscopic objects of comparable
size to the current Kuiper belt objects, formed by coagulation), and
`disk particles' (golf-ball sized planetesimals that constitute the
bulk of the disk's mass and which have a very intense collision rate
and damping).  The KBOs are not massive enough to be affected by
dynamical friction, but are big enough not to be damped by collisions
with the disk-particles.

As the planets grow to their final sizes, C06 estimate that the KBOs
can be scattered by the growing planets to orbits with eccentricities
and inclinations of order of 0.2.  These, they argue, become the hot
population, which has observed eccentricities and inclinations
comparable to these values.  After \Red{all but two of the} original
planets are removed by the dynamical instability and the ice giants
evolve onto their current orbits inside of $30\,$AU, C06 suggest that
there is still a population of \Red{very small} disk particles between
40 and $47\,$AU.  With the planets gone, this disk can become
dynamically cold enough to allow large objects to grow in it,
producing a second generation of KBOs on low-eccentricity and
low-inclination orbits.  These objects should be identified, in C06's
scenario, with the observed cold population of the Kuiper belt.

The mass of the disk between 40 and $47\,$AU, however, should retain
on order of half of its original mass, namely about 20~$M_\oplus$
according to the surface density assumed in C06's Equation~(13). How
this total mass was lost and how the cold population acquired its
current, non-negligible eccentricity excitation, are not really
explained by C06. The authors limit themselves to \Red{a discussion
  of} the radial migration of Neptune, after the circularization of
its orbit, \Red{to create the resonant populations (Malhotra~1995 Hahn
  \& Malhotra 1999).  This, in principal, might excite the cold
  classical belt,} although \Red{C06} admit that it is not at all
obvious how planet migration would proceed in a highly collisional
disk.  As for the mass depletion, \Red{collisional grinding is the
  only mechanism that makes sense at this point in time.}  However, it
is not clear (at least to us) why collisional grinding would become so
effective at this late stage while it was negligible during the planet
formation and removal phases, when the relative velocities \Red{were}
much higher.

At this point we want to emphasize that, although the ideas presented
in G04 and C06 are new and intriguing, the papers do not present any
actual models.  Most of the arguments are based on order-of-magnitude
equations where factors of 2 and $\sqrt{3}$ are dropped and
approximate time-scales are set equal to one another in order to
determine zeroth-order steady state solutions.  In addition,
simplifications are made to make the problem tractable analytically,
like, for example, at any given instant all of the disk particles have
the same size. Another example of a simplification is that the rate of
change of the velocity dispersion of the disk particles due to
collisions is simply set to the particle-in-the-box collision rate ---
the physics of the collisions are ignored.  While making such
approximations is reasonable when first exploring a problem and
determining whether it could possibly work, numerical experiments are
really required in order to determine whether the process does indeed
act as the analytic expressions predict.

Finally, many of the steps in these scenarios are not justified.  Of
particular interest to us is the stage when, according to G04,
\Red{all but two (i$.$e$.$ three of five in the nominal case) of the
  original} ice giants are removed from the system via a gravitational
instability.  The papers present order-of-magnitude equations that
argue that such an instability would occur, but the authors are forced
to speculate about the outcome of this event.

Indeed, we suspect, based on our experiences, that G04's expectations
about the removal of the ice giants are naive.  In particular, Levison
et al$.$~(1998) followed the dynamical evolution of a series of
fictitious giant planet systems during a global instability.  They
found that during the phase when planets are scattering off of one
another, the planetary system spreads to large heliocentric distances,
and, while planets can be removed by encounters, the outermost planet
is the most like to survive.  Similarly, Morbidelli et al$.$~(2002)
studied systems of planetary embryos of various masses originally in
the Kuiper belt and found that in all cases the embryos spread and
some survived at large heliocentric distances.  From these works we
might expect that G04's instability would lead to an ice giant at
large heliocentric distances (but still within the observation
limits), rather than having a planetary system that ends at $30\,$AU
with a disk of small particles beyond.  Granted, the simulations in
both Levison et al$.$~(1998) and Morbidelli et al$.$~(2002) did not
include a disk of highly-damped particles that can significantly
affect the evolution of the planets, so new simulations are needed to
confirm or dismiss the G04/C06 scenario.  In this paper we perform
such simulations.  We are required to develop a new numerical
integration scheme to account for the collisional damping of the
particle disk.  This scheme is detailed and tested in the next
section.

\medskip
\medskip

\section{The Code}
\label{sec_code}

In this section we describe, in detail, the code that we constructed
to test the C06 scenario.  Before we can proceed, however, we must
first discuss what physics we need to include in the models.  As we
described above, our motivation is to determine how a system
containing a number of ice-giant planets embedded in a disk of
collisionally damped particles dynamically evolves with time.  Our
plan is to reproduce the systems envisioned by G04 and C06 as closely
as possible rather than create the most realistic models that we can.
Thus, we purposely adopt some of the same assumptions employed by G04.
For example, although G04 invoke a collisional cascade to set up the
systems that they study, their formalism assumes that the disk
particles all have the same size and ignore the effects of
fragmentation and coagulation.  We make the same assumption.

In addition, although G04 invokes a collisional cascade to grind
kilometer-sized planetesimals to submeter-sized disk particles, they
implicitly assume that the timescale to change particle size is short
compared to any of the dynamical timescales in the problem.  Thus,
their analytic representation assumes that the radius of the disk
particles, $s$, is fixed.  They determine which $s$ to use by arguing
that disk particles will grind themselves down until the timescale for
the embryos to excite their orbits is equal to the collisional damping
time (which is a function of $s$).  Then $s$ is held constant.  We,
therefore, hold $s$ constant as well.  In addition, G04 does not
include the effects of gas drag in their main derivations, we again
follow their lead in this regard.

Our code is based on SyMBA (Duncan et al. 1998, Levison \& Duncan
2000). SyMBA is a symplectic algorithm that has the desirable
properties of the sophisticated and highly efficient numerical
algorithm known as Wisdom-Holman Map (WHM, Wisdom \& Holman 1991) and
that, in addition, can handle close encounters (Duncan et al$.$~1998).
This technique is based on a variant of the standard WHM, but it
handles close encounters by employing a multiple time step technique
introduced by Skeel \& Biesiadecki~(1994). When bodies are well
separated, the algorithm has the speed of the WHM method, and whenever
two bodies suffer a mutual encounter, the time step for the relevant
bodies is recursively subdivided.

Although SyMBA represented a significant advancement to the
state-of-art of integrating orbits, it suffers from a basic and
serious limitation. At each time step of the integration, it is
necessary to calculate the mutual gravitational forces between all
bodies in the simulation. If there are $N$ bodies, one therefore
requires $N^2$ force calculations per time step, because every object
needs to react to the gravitational force of every other body. Thus,
even with fast clusters of workstations, we are computationally limited
to integrating systems where the total number of bodies of the order
of a few thousand.

Yet, in order to follow both the dynamical and collisional evolution
of the numerous small bodies present during the G04's scenario, we
need to implement a way to follow the behavior of roughly $10^{26-29}$
particles. This clearly is beyond the capabilities of direct orbit
integrators. Only statistical methods can handle this number of
objects.  In the following, we describe our approach to this problem.

As described above, the systems that C06 envisions have three classes
of particles: the planetary embryos, the KBOs, and the disk particles.
Each class has its unique dynamical characteristics.  The embryos are
few in number and their dynamics are not directly effected by
collisional damping.  Thus, in our new code, which we call {\it
  SyMBA\_COL}, they can be followed directly in the standard $N$-body
part of SyMBA.  The KBOs are not dynamically important to the system
from either a dynamical or collisional point of view.  Since we are
more concerned here with the final location of the ice giants than the
dynamical state of the Kuiper belt, we ignore this population.
Finally, we need to include a very large population of submeter-sized
particles that both dynamically interact with the rest of the system
and collisionally interact with each other.

Thus, we have added a new class of particle to SyMBA which we call a
{\it tracer} particle.  Each tracer is intended to represent a large
number of disk particles on roughly the same orbit as one another.
Each tracer is characterized by three numbers: the physical radius
$s$, the bulk density $\rho_b$, and the total mass of the disk
particles represented by the tracer, $m_{\rm tr}$.  For example, the
runs presented below, we set $s=1\,$cm or $1\,$m and $\rho_b = 1\,{\rm
  g\,cm^{-3}}$.  In addition, we typically want to represent a
$\sim\!75\,\mearth$ disk with $\sim\!3000$ tracer particles, meaning
that $m_{\rm tr}\!\sim\!0.025\,\mearth$ (although the exact numbers
vary from run to run).

The first issue we needed to address when constructing SyMBA\_COL, was
to determine an algorithm that correctly handles the gravitational
interaction between the embryos and the disk particles.  Since there
are only a few embryos and they are relatively large, the acceleration
of the tracers due to the embryos can be determined using the normal
$N$-body part of SyMBA.  It is less obvious, however, whether the
gravitational effect of the disk particles on the embryos can also be
effectively simulated using the normal $N$-body part of SyMBA,
i$.$e$.$ using the forces directly exerted on the embryos by the
tracers. To argue this position, let us point out that the
gravitational effect of the disk particles on the embryos is well
approximated by the {\it dynamical friction} formalism, which,
assuming a Maxwellian velocity distribution, can be written as
(Chandrasekhar 1943; also see Binney \& Tremaine~1987):
\begin{equation}
{d\vec{w}\over dt} \propto {(m_{\rm em}+m_{\rm dp}) \rho_{\rm disk}
  \over w^3} \left[{\rm erf}(X) - \frac{2X}{\sqrt{\pi}}
  e^{-X^2}\right] \vec{w},
\label{eq_df}
\end{equation}
where $X\!\equiv\!w/(\sqrt{2}u$), $w$ is the velocity of the embryo,
$u$ is the velocity dispersion of the disk particles, $m_{\rm dp}$ is
that mass of an individual disk particle, `erf' is the error function,
and $\rho_{\rm disk}$ is the background volume density of the disk.
So, if $m_{\rm em} \gg m_{\rm dp}$, the acceleration of the embryos
due to the disk is independent on the mass of individual disk
particles.  Thus, although $m_{\rm tr}\!\gg\!m_{\rm dp}$, the direct
acceleration of the embryos due to the tracers is roughly the same as
if we had each individual disk particle in the simulation as long as
$m_{\rm tr}\!\ll\!m_{\rm em}$.  Therefore, in general, we can employ
the standard $N$-body part of the SyMBA to calculate the gravitational
effects of the embryos and the disk particles on each other.  We
return to the issue of how big $m_{\rm em}/m_{\rm tr}$ needs to be in
the next section.

All that is left is to consider the effects of the disk particles on
each other.  There are two effects that must be included: collisional
damping and self-gravity.  We handle the former through Monte Carlo
techniques.  The first step in our collisional algorithm is to divide
the Solar System into a series of logarithmically spaced annular rings
that, in the simulations performed here, stretched from $10\,$AU to
$60\,$AU.  As we describe more below, the logarithmic spacing is
employed by the self-gravity algorithm.  In all, we divided space into
$N_{\rm ring}\!=\!1000$ such rings in our production runs (although in
some of our tests, we used $N_{\rm ring}\!=\!10,000$ rings).  We use
these rings to statistically keep track of the state of the disk
particles.  In particular, as the simulation progresses, we keep track
of the tracer particles moving through the ring and from this
calculate: 1) the total mass of disk particles in that ring ($M_a$),
and 2) the vertical velocity dispersion of the disk particles, $u_z$.
These values are recalculated every $\tau_{\rm update}$, by moving
each tracer along its Kepler orbit in the barycentric frame and adding
its contribution to each ring as it passes through.  We set $\tau_{\rm
  update} = 200\,$yr.  In addition, as the tracers orbit during the
simulation, we keep a running list of their velocities and longitudes
as they pass through each individual ring.  Entries are dropped from
this list if they are older than $\tau_{\rm update}$.

At each timestep in the simulation, we evaluate the probability, $p$,
that each tracer particle will suffer a collision with another disk
particle based on the particle-in-a-box approximation.  In particular,
$p \equiv n\, (4\pi s^2)\, w\, dt$, where $n$ is the local number
density of the disk particles, $w$ is the velocity of the tracer
relative to the mean velocity of the disk, and $dt$ is the timestep.
It is important to note that $n$ is not the number density of tracers,
but the number density of the disk assuming that all disk particles
had a radius of $s$.  Therefore, $p$ does not carry any information
about the mass or number of tracers, but is only dependent on the
surface density of the disk, its vertical velocity dispersion, and
$s$.  In addition, the analytic derivations usually assume that
$w\!=\!u$, however, here we use the true velocity of the individual
tracer.

We assume that
\begin{equation} 
n(z) = \Biggm\{ \begin{array}{cc} \left(1-\exp{[-1]}\right)\,n_0 & ~~~~{\rm if~} z<z_0 \\ 
                                n_0\exp{[-z/z_0]} &  ~~~~{\rm if~} z \ge z_0,  
\end{array}
\end{equation} 
where, $z$ is the distance above the disk mid-plane, $n_0 = M_a /
2\,(\frac{4}{3}\pi s^3 \rho_b)\,z_0\,A_a$, $A_a$ is the area of the
ring, and $z_0$ is the scale height of the disk.  In particular, $z_0
= u_z/\Omega$, where $\Omega$ is the orbital frequency.  We hold
$n(z)$ constant for $z<z_0$ to help correct for the fact that $dt$ is
finite.  If we did not hold $n$ constant there would be a danger that
we would underestimate the collision rates because particles would
jump over the high density mid-plane as they orbit.  The price to pay
for this is a discontinuity in $n(z)$ at $z_0$, but it is a price, we
believe, that is wise to pay.

Once we have determined $p$, we generate a random number between 0 and
1, and if $p$ is larger than this number, we declare that the tracer
has suffered a collision with another disk particle.  Now, we need to
determine the velocity of the impactor, and again we turn to our
rings.  As we stated above, as the tracers orbit during the
simulation, we keep a running list of particles that had passed
through each ring, keeping tract of their individual velocities.  The
impactor is assumed to have the same location as the target, but its
velocity is chosen from this running list appropriately rotated
assuming cylindrical symmetry.  We also assume that two particles
bounce off of one another (as in G04), but that the coefficient of
restitution is very small.  The end result is that we change the
velocity of our target tracer to be the mean of its original velocity
and that of the impactor.

We have to spend a little time discussing how we decide which velocity
in our running list to choose because if this is done incorrectly it
leads to a subtle error in the results.  In an early version of the
code we simply chose the velocity at random.  This is the same as
assuming that the disk is axisymmetric, and thus collisions try to
force particles onto circular orbits.  However, in a situation where
the disk is interacting with a massive planet on an eccentric orbit,
the natural state of the disk is for its particles to evolve onto
eccentric orbits whose longitude of perihelion, $\tilde\omega$, is the
same as the planet's, and whose eccentricity, $e$, is a function of
the eccentricity of the planet and the distance from the planet.  In
essence, this situation should produce an eccentric ring.
Experimenting with our original code showed that in situations where
the ring is massive, the assumption of axisymmetry causes the planet
to migrate away from the disk at an unphysically large rate, as
collisions try to force particles onto circular orbits while the
planet tries to excite their eccentricities.

Thus, we found that our code needs to be able to support eccentric
rings.  We found we can accomplish this by modifying the method we use
for choosing a velocity for the impactor.  In particular, in addition
to storing a particle's velocity in the running list, we also keep
tract of its true longitude, $\lambda$.  We choose from the running
list the velocity of the object that has the $\lambda$ closest to that
of the target tracer.  In this way, asymmetries can be supported by
the code.  We show an example of this in the test section below.

Through experimentation we also found that we need to include
self-gravity between the disk particles, at least crudely.  In an
early version of the code we did not include this effect and found
that under certain conditions there was an unphysical migration of
planetary embryos.  In particular, in situations where disk particles
became trapped in a MMR with an embryo (particularly the 1:1 MMR) the
embryo was incorrectly pushed around by the disk particles if there
was a large number of additional disk particles in the system.  This
did not occur if the particle self-gravity was included.

There are too many tracers in our simulations to include self-gravity
directly.  Thus, we employ a technique originally developed for the
study of disk galaxies, known as the particle-mesh (PM) method
(Miller~1978).  In what follows, we use the formalism from Binney \&
Tremaine~(1987).  We first define a modified polar coordinate system
$\Red{{\rm u}}\!\equiv\ln{\varrho}$ and $\phi$, where $\varrho$ and
$\phi$ are the normal polar coordinates, and define a {\it reduced}
potential, $V(\Red{{\rm u}},\phi) = e^{\Red{{\rm u}}/2} \Phi\left[\varrho(\Red{{\rm u}}),\phi\right]$ and
a {\it reduced} surface density $S(\Red{{\rm u}},\phi) =
e^{\Red{{\rm u}}/2}\sigma\left[\varrho(\Red{{\rm u}}),\phi\right]$ such that:
\begin{equation}
V(\Red{{\rm u}},\phi) = -{G\over\sqrt{2}} \int\limits_{-\infty}^{\infty}
\int\limits_{0}^{2\pi} {S(\Red{{\rm u}}',\phi') d\phi' \over \sqrt{\cosh{(\Red{{\rm u}}-\Red{{\rm u}}')} -
\cos{(\phi-\phi')}}} d\Red{{\rm u}}'
\end{equation}
If we break the disk into cells this becomes: 
\begin{equation}
\label{eq_Vlm}
V_{lm} \approx
\sum\limits_{l'} \sum\limits_{m'} {\cal G}(l'-l, m'-m) {\cal
  M}_{l'm'}
\end{equation}
where ${\cal M}_{lm} = \int\int_{{\rm cell}(l.m)} S\,d\Red{{\rm u}}\,d\phi$ and
${\cal G}$ is the Green's function:
\begin{equation}
  \label{eq_gr}
  {\cal G}(l'-l, m'-m) = -{G \over \sqrt{2\left(\cosh{(\Red{{\rm u}}_{l'}-\Red{{\rm u}}_l)} -
      \cos{(\phi_{m'}-\phi_{m})}\right)}},
\end{equation}
when $l\neq l'$ and $m\neq m'$, and 
\begin{equation}
{\cal G}(0,0) =
-2G\left[{1\over{\Delta\phi}}\sinh^{-1}{\left(\Delta\phi\over\Delta
      \Red{{\rm u}}\right)} + {1\over{\Delta \Red{{\rm u}}}}\sinh^{-1}{\left(\Delta
      \Red{{\rm u}}\over\Delta \phi\right)}\right],
\end{equation}
where ${\Delta \Red{{\rm u}}}$ and $\Delta\phi$ are the grid spacings.

For this algorithm we found that it is best to assume that the disk is
axisymmetric, so Equation~\ref{eq_Vlm} becomes
\begin{equation}
V_{lm} \approx \sum\limits_{l'}
\sum\limits_{m'} {\cal G}(l'-l, m'-m) {\Delta\phi\over2\pi} {\cal
  M}_{l'} = \sum\limits_{l'} {\Delta\phi\over2\pi} {\cal M}_{l'}
\sum\limits_{m'} {\cal G}(l'-l, m'-m)
\end{equation}
\begin{equation}
\label{eq_Vl}
V_l \approx \sum\limits_{l'}
{\Delta\phi\over2\pi}
{\cal M}_{l'} \tilde{\cal G}(l'-l).
\end{equation}
Note that Equation~\ref{eq_Vl} is one dimensional, and thus it only
supplies us with a radial force.  The tangential and vertical forces
are assumed to be zero.  We made this assumption due to the small
number of tracers in our system.  However, a simple radial force is
adequate for our purposes.  

Also, the form of Equation~\ref{eq_Vl} allows us to use the rings
already constructed for the collisional algorithm.  All we need is that
relationship between ${\cal M}_{l'}$ and the total amount of mass in
ring, $M_a$.  We find that 
\begin{equation}
{\cal M}_{l'} = {2 M_a\over\left(a_{l2}^2
    - a_{l1}^2\right)} a_l^{3/2} \left[\ln(a_{l2}) -
  \ln(a_{l1})\right], 
\end{equation}
where $a_{l2}$, $a_{l1}$, and $a_l$ are the outer edge, inner edge,
and radial center of ring $l$.

So, Equation~\ref{eq_Vl} gives us the reduced potential at the center
of ring $l$ and thus the true potential can be found ($\Phi =
e^{-\Red{{\rm u}}/2}V$).  To calculate the radial acceleration at any
location, we employ a cubic spline interpolation scheme.  Finally, the
acceleration of a particle is calculated by numerically
differentiating this interpolation.

\medskip
\medskip

\subsection{Tests}
\label{sec_tests}

In this subsection we describe some of the tests that we performed on
SyMBA\_COL. 

\medskip


\noindent \underline{\it An Isolated Ring:}  In this test, we
\Red{studied} the behavior of a disk of particles initially on
eccentric orbits as collisions damp their relative velocities.  In
particular, we evolved a system containing the Sun and 1000 tracers,
which were uniformly spread in semi-major axis from 30 to $35\,$AU.
Initially, the tracers were given a Raleigh distribution with an RMS
eccentricity equal to 0.1, and an RMS $\sin{(i)}$ equal to $0.05$.
The total mass of the ring was $10\,\mearth$ and we set $s\!=\!1\,$cm.

\begin{figure}[h!]
\vglue 3.0truein
\includegraphics{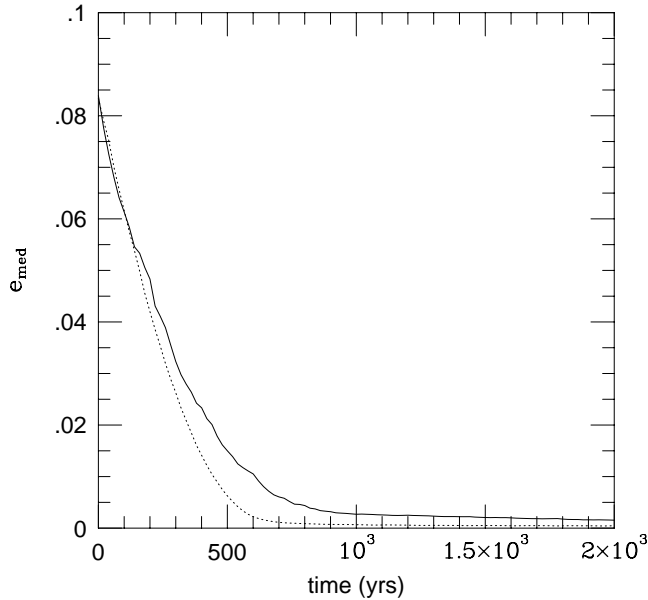}
\caption{\footnotesize \label{fig_t1f_e}
  {The temporal evolution of the collisionally active ring of
    particles used as the first test of our new code.  In particular,
    we plot the median eccentricity ($e_{med}$).  The solid and dotted
    curves refer to the normal and HIRES runs, respectively.}}
\end{figure}

The solid curve in Figure~\ref{fig_t1f_e} shows the temporal evolution
of the eccentricities in the above system. As the system evolved, the
ring collapsed (i$.$e$.$ eccentricities and inclinations dropped) as
the collisions damped out random velocities.  During this process, we
found that a small fraction of the particles were left behind because
as the ring collapsed these particles found themselves in regions of
space where there were no other particles.  For example, if a particle
had a relatively large initial inclination and happened not to have
suffered a collision early on, then it can be left behind on a large
inclination orbit because it finds itself traveling above and below
the collapsing disk most of the time.  Eventually, it will hit another
particle because it penetrates the disk, but this can take a long
time.  The end result of this process is that during the collapse,
there is always a high velocity tail to the eccentricity and
inclination distributions.  To correct for this, we plot the median
eccentricity rather than the more standard RMS eccentricity in
Figure~\ref{fig_t1f_e}.  In this system, we find that the e-folding
damping time of the eccentricity is $320$ years.
 
In order to test whether our code has converged we performed a second
experiment.  Recall that in the first run, we used 1000 tracers,
$N_{\rm ring}\!=\!1000$, and $\tau_{\rm update}\!=\!200\,$yr.  This
produced the solid curve in the figure.  The dotted curve shows the
results for a high resolution run where $N_{\rm ring}\!=\!10,000$,
$\tau_{\rm update}\!=\!20\,$yr, and we used $10,000$ tracer particles.
\Red{Although there are some differences between the two curves, the
  basic evolution of the systems, including their e-folding damping
  times, are the same.  Thus, we feel that the code has converged well
  enough that we can employ the lower resolution.}

We now need to calculate what G04's development would predict.  G04a's
Equation~(50) states that
\begin{equation}
\label{eq_udot}
{1\over\tau_{\rm col}} \equiv - {1\over u}{du\over dt} \sim  \Omega
  {\sigma \over \rho_b s}.
\end{equation}
Plugging in the appropriate values for this test, we find that G04
predicts that the collisional damping time, $\tau_{\rm col}$, should
be $\sim\!110\,$yr, which is about a factor of 3 shorter than we
observed.  However, we believe that the agreement is reasonably good
because G04's derivations were intended to be order-of-magnitude in
nature and that factors of a few were typically dropped.

\medskip

\begin{figure}[h!]
\vglue 2.4truein
\includegraphics{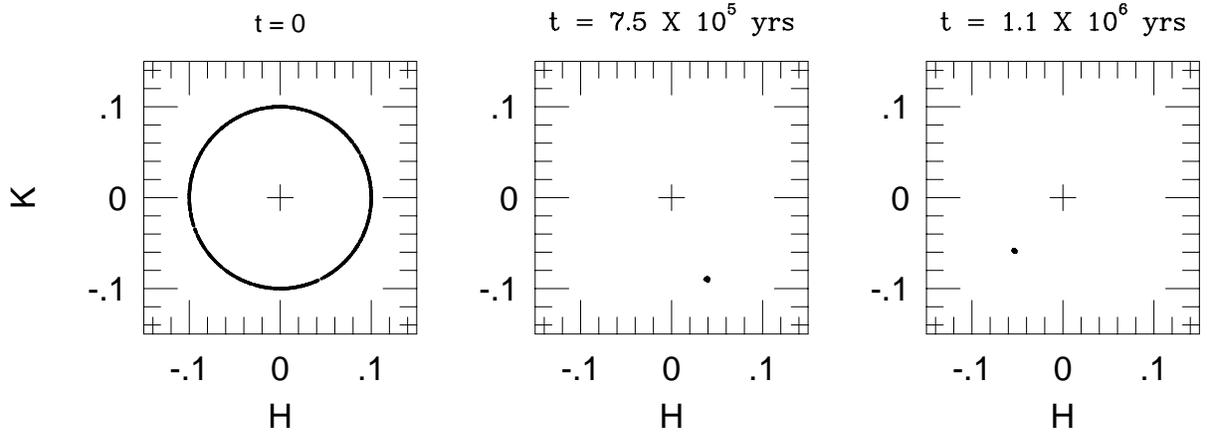}
\caption{\footnotesize \label{fig_t8_HK}
   {Three snapshots of the evolution of a ring of collisionally
    active particles under the gravitational influence of a
    Saturn-mass planet with $e=0.2$.  The ring has a semi-major axis
    1.72 times larger than the planet.  The particles, which were
    initially uniform in $\tilde\omega$
    ($H\!\equiv\!e\cos{(\tilde\omega)}$ and
    $K\!\equiv\!e\sin{(\tilde\omega)}$), clump in $\tilde\omega$ and
    thus form an eccentric ring.}}
\end{figure}

\noindent \underline{\it \Red{A} Narrow Ring with a Giant Planet:}  As
discussed above, we \Red{found} that our code must be able to support
an eccentric ring if the dynamics demand it, and we described our
methods for doing so.  Here we present a test of this ability.  In
particular, we studied the behavior of a $10\,\mearth$ narrow ring of
collisional particles under the gravitational influence of Saturn.  In
order to enhance the eccentricity forcing of the ring, we set Saturn's
$e=0.2$.  The semi-major axes ($a$) of the disk particles was spread
from $1.70 < a/a_{\rm Sat} < 1.72$ where $a_{\rm Sat}$ is Saturn's
semi-major axis.  The particles had an initial eccentricity of 0.1.
Of importance here, the initial $\tilde\omega$ was randomly chosen
from the range of 0 and $2\pi$.  Thus, if we define two new variables,
$H\!\equiv\!e\cos{(\tilde\omega)}$ and
$K\!\equiv\!e\sin{(\tilde\omega)}$, then at $t\!=\!0$ the tracer
particles fall along a circle of radius $e$ in $H$--$K$ space
(Figure~\ref{fig_t8_HK}).

We find that as the system settles down, the particles evolve to a
point in $H$--$K$ space (Figure~\ref{fig_t8_HK}).  And since this
point does not sit at the origin, the particles have the same $e$ and
$\tilde\omega$, i$.$e$.$ they form an eccentric ring.  Unfortunately,
we know of no analytic theory describing the behavior of this ring.
However, we can take a clue from the secular theory of the response of
a massless test particle to an eccentric planet (Brouwer \& Clemence
1961, see also Murray \& Dermott~2000), which predicts that the
`forced' eccentricity of the particle is:
\begin{equation}
\label{eq_eforce}
e_{\rm forced} = {b_{3/2}^{(2)}\left(a_{\rm Sat}/a\right) \over
  b_{3/2}^{(1)}\left(a_{\rm Sat}/a\right)}~e_{\rm Sat}, 
\end{equation}
where $b_{3/2}^{(i)}$ are the Laplace coefficients. Plugging in the
appropriate values, we find $e_{\rm forced}\!=\!0.14$.  This should be
an upper limit to the actual eccentricity of the ring because the
collisions within the disk, which are not included in
Equation~\ref{eq_eforce}, should decrease eccentricity.  We find the
ring has an eccentricity of $\sim\!0.1$, although the exact values
changes over time.  Thus, the ring seems to behaving reasonably.
Interestingly, we also find that this ring's precession is negative,
which implies that self-gravity is important to its dynamics, which
again is reasonable.


\medskip


\noindent \underline{\it A System of Growing Embryos in the
  Dispersion-Dominated Regime}: As G04 explained in detail, a system
consisting of growing embryos in a sea of disk particles will be in
one of two possible modes. If the velocity dispersion of the disk
particles is large enough that the scale height of the planetesimal
disk exceeds the radius of the embryo's Hill's sphere, $r_H = a \left(
  m_{\rm em}/3 M_\odot\right )^{1/3}$, then the disk behaves as if it
is fully three dimensional.  This occurs if $u \gapprox \Omega r_H$.
This situation is \Red{referred} to as the {\it dispersion-dominated}
regime.  However, if collisions damp planetesimal random velocities
strongly enough, $u$ can get much smaller than $\Omega r_H$ and the
system enters the so-called {\it shear-dominated} regime. In this
mode, growth proceeds in a qualitatively different way, and can be
much more rapid than dispersion-dominated growth (also see
Rafikov~2004).  In the extreme, the velocity dispersion can be so
small that the entire vertical column of the planetesimal disk is
within the protoplanet's Hill's sphere, thus making accretion a
two-dimensional process.

Given the different nature of these two regimes, we test each of them
separately.  We \Red{begin} with the dispersion-dominated regime.
This test \Red{starts} with a population of ten $5000\,$km radii
embryos spread in semi-major axes between 25 and $35\,$AU.  This
implies that $\Sigma\!=\!4.6\times\10^{-4}\,\mearth/{\rm
  AU}^2\!=\!0.012\,{\rm gm/cm}^2$.  The initial eccentricities of the
embryos were set to 0, but the inclinations were given a Raleigh
distribution with an RMS $\sin{(i)}$ equal to $1.5\times10^{-4}$ to
insure that the embryo-embryo encounters can excite inclinations.

\begin{figure}[h!]
\vglue 4.2truein
\includegraphics{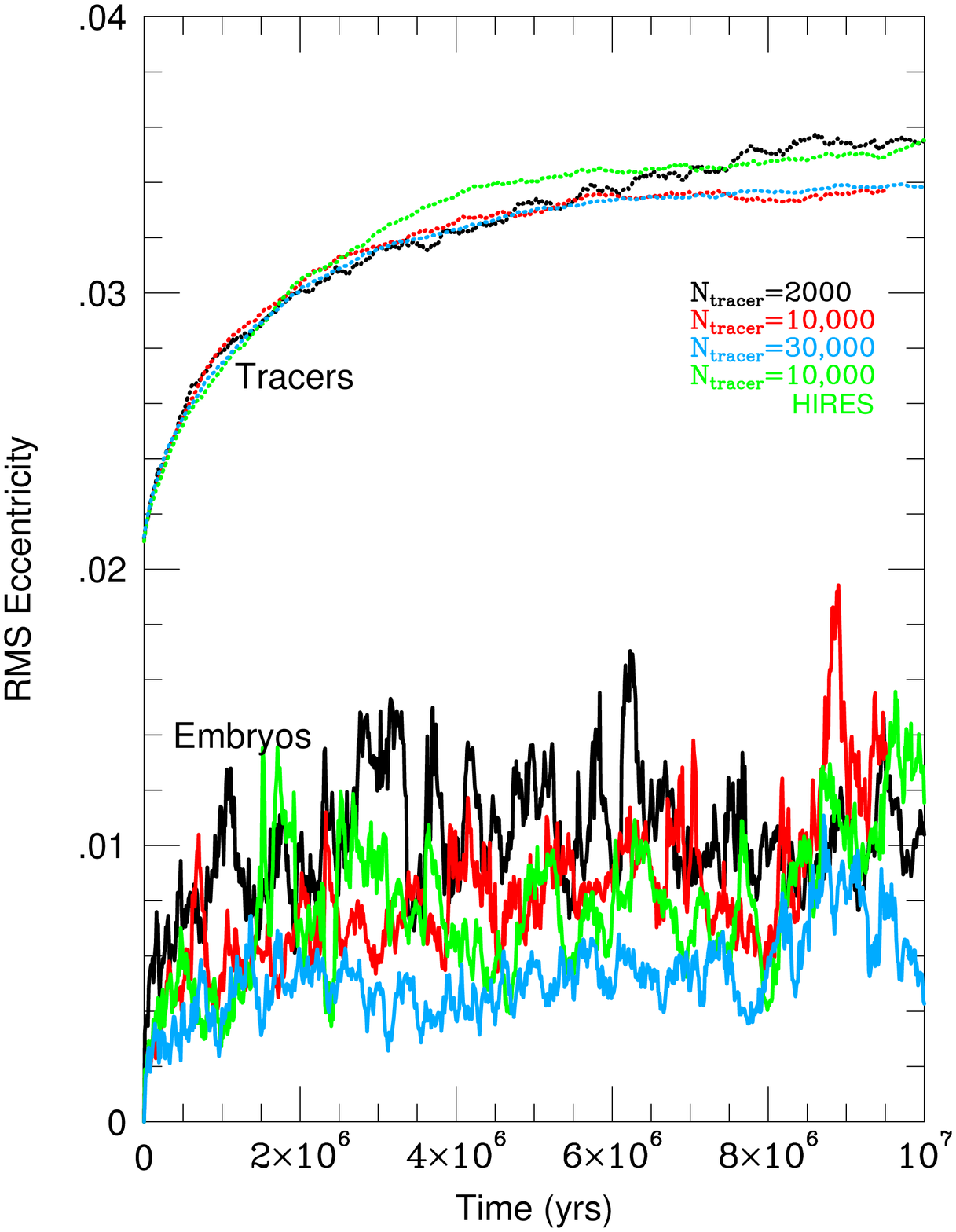}
\includegraphics{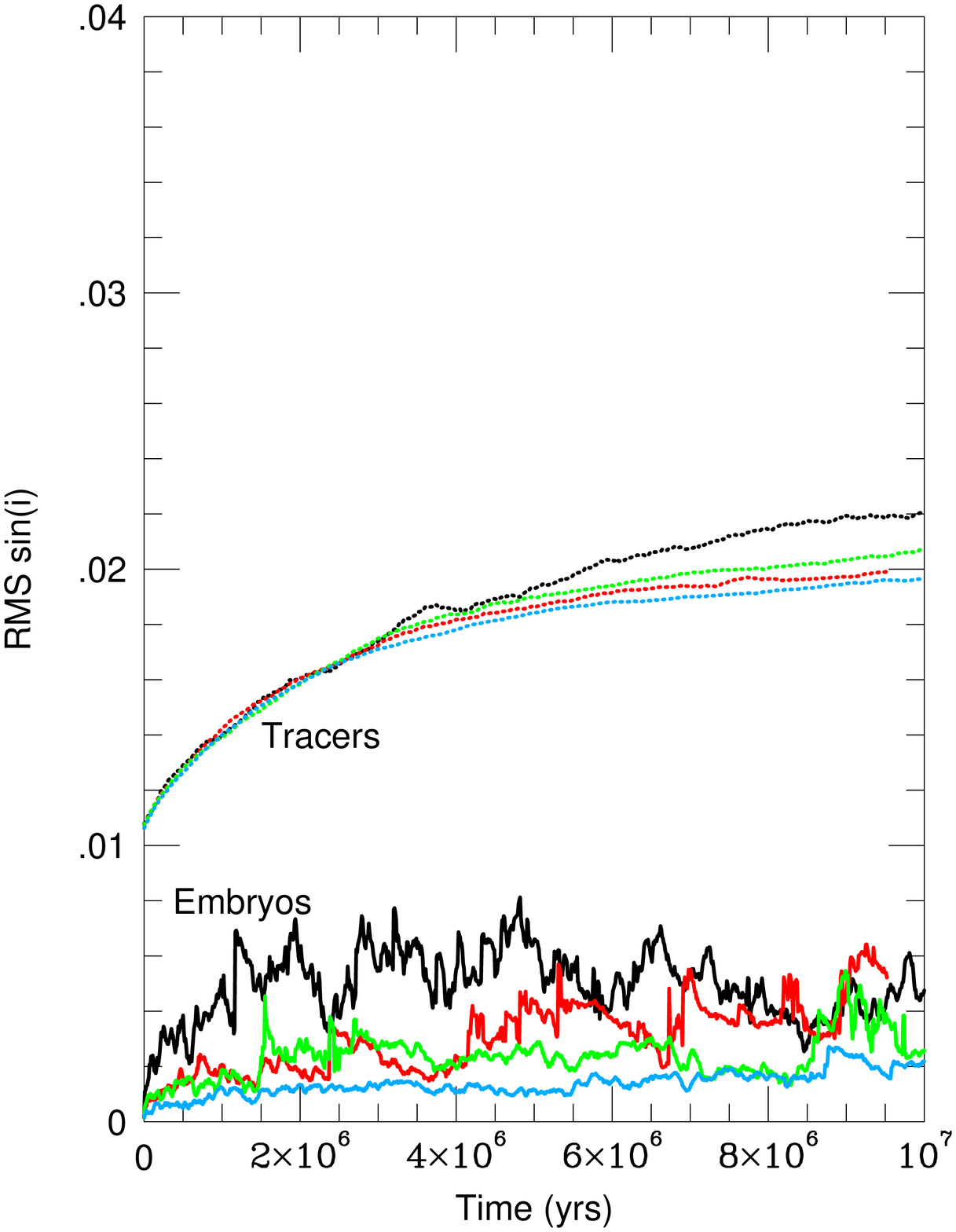}
\caption{\footnotesize \label{fig_t8f_ei}
  {The temporal evolution of the RMS eccentricity (left) and the RMS
    inclination (right) of a system of embryos embedded in a disk of
    small objects.  The disk particles are collisionally active. The
    solid curves shows the behavior of the embryos, while the dotted
    curves shows that of the disk particles.  Different colors refer
    to different disk resolutions, see text and legend for a
    description.}}
\end{figure}

The disk was spread between 20 and $40\,$AU and was designed so that
$\sigma\!=\!1.5 \10^{-3}\,\mearth/{\rm AU}^2$.  We set $s\!=\!100\,$m.
The disk was represented by at least 2000 tracers, where the
eccentricities and inclinations were chosen from Raleigh distribution
with an RMS eccentricity and $2\sin{(i)}$ equal to $0.025$.
Figure~\ref{fig_t8f_ei} shows the temporal evolution of the
eccentricities and inclinations of both the embryos (solid curves) and
the disk particles (dotted curves) as observed in several runs where
we varied $N_{\rm tracer}$ and $N_{\rm rings}$.

G04 predicts that the eccentricities of both the embryos and disk
particles would increase until a steady state is reached.  This steady
state is caused by a balance in the heating and cooling processes in
both populations.  The disk particles are being excited by the
embryos, while they are being damped by collisions.  At the same time,
the embryos are exciting each other, while they are being damped due
to dynamical friction with the disk particles.  This steady state is
clearly seen in our simulations (Figure~\ref{fig_t8f_ei}).

Indeed, G04's analysis allows us to predict what the steady state
eccentricities should be.  In particular, in the regime of interest,
G04 finds (Equation~76) that for the disk particles
\begin{equation}
u \sim v_{\rm esc} \left({\Sigma\over \sigma} {s\over R} \right) ^ {1\over 4},
\end{equation}
where $v_{\rm esc}$ is the escape velocity of the embryos.  Plugging
in the above values, we find that $u=0.04\,{\rm AU\, yr^{-1}}$, which
corresponds to an RMS eccentricity of $\sim\!0.04$ (the circular
velocity at $30\,$AU is $1.1\,{\rm AU\, yr^{-1}}$ so that
$e\!\sim\!u$).  This is in excellent agreement with our simulation.
For the embryos, if $v$ is defined to be their velocity dispersion,
then recently Chiang \& Lithwick (2005, their Equation 45) showed that
\begin{equation}
{v\over u} \sim \left({\Sigma\over 8\sigma}\right) ^ {1\over 2},
\end{equation}
which is $0.2$ in our test problem.  We observe this ratio to be
roughly $0.3$.  We believe that our simulations are reproducing the
G04's predictions fairly well given the order-of-magnitude nature of
G04's derivations.

As in our previous test, we must consider whether our code has high
enough resolution.  To investigate this issue, we performed four
simulations where the only difference was the resolution of the disk.
The black, red, and blue curves in Figure~\ref{fig_t8f_ei} show the
results for an increasing number of tracer particles (2000, $10,000$,
and $30,000$, respectively).  In all these cases $N_{\rm
  ring}\!=\!1000$.  In addition, the green curves (labeled `HIRES')
show the results for a simulation with $10,000$ tracer particles, but
where $N_{\rm ring}\!=\!10,000$.  In all cases the behavior of the
systems are very similar.  \Red{Indeed, the code has converged
  adequately enough, particularly when $N_{\rm
    tracers}\!\gtrsim\!10,000$.  As we discuss below, such resolutions
  are required in order to handle the shear-dominated regime.}

We must also consider conserved quantities when testing our code.
Unfortunately, collisions do not conserve energy, and so the only
conserved quantity is the angular momentum vector of the system.  In
the run presented in Figure~\ref{fig_t8f_ei}, we find that angular
momentum is conserved to 1 part in $10^4$ over $100\,$Myr, which is
satisfactory.

Finally, G04 predict that the embryos will grow as they accrete the
disk particles.  In particular, their Equation~77 predicts that
\begin{equation}
{dR\over dt} \sim {\Omega \sigma \over \rho_b} \left(
{\Sigma\over \sigma}  {s\over R} \right)^{-{1\over 2}}.
\label{eq_dr}
\end{equation}
For the parameters in this test, this predicts
$dR/dt\!\sim\!6\10^{-6}\,{\rm km/yr}$.  In the simulation, all embryos
started with $R\!=\!5000\,$km.  At the end of $10^7$ years,
Equation~\ref{eq_dr} forecasts that the embryos should grow
$\sim\!50\,$km.  We find at the end of the simulation that the average
embryo radius is $5038\,$km. Again, the agreement is good.

\medskip


\noindent \underline{\it A System of Growing Embryos in the
  Shear-Dominated Regime}: As a final test of our code, we study the
behavior of a system containing \Red{three} embryos with a mass of
$0.17\mearth$ embedded in a $7\mearth$ disk of planetesimals spread
from 27 to $33\,$AU.  The radius of the disk particles was set to
$5\,$cm.  This problem was designed so that the system should be in
the shear-dominated regime.

\begin{figure}[h!]
\vglue 5.0truein
\includegraphics{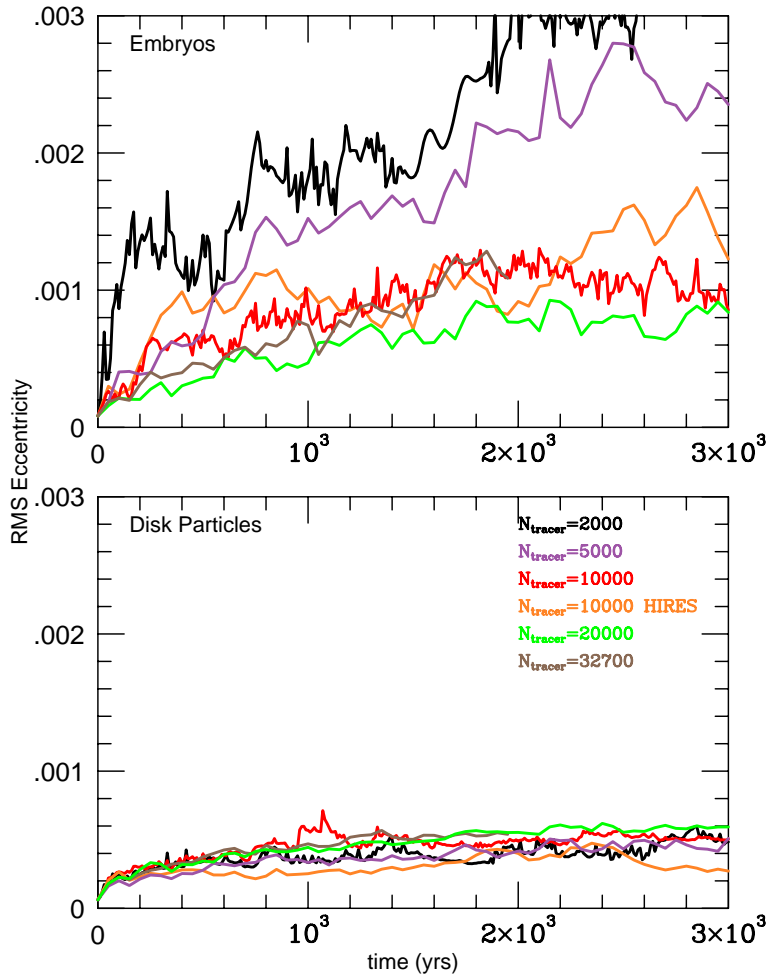}
\caption{\footnotesize \label{fig_t5f_e}
  {The temporal evolution of the RMS eccentricity of a system of
    embryos embedded in a disk of small objects in the shear-dominated
    regime.  The top panel shows the behavior of the embryos, while
    the bottom shows that of the disk particles.  Different colors
    refer to different disk resolutions, see text and legend for a
    description.}}
\end{figure}

The results from several simulations with different disk resolutions
is presented in Figure~\ref{fig_t5f_e}.  We start with a discussion of
code convergence.  We preformed simulations with $N_{\rm tracer}$
between 2000 and 32,700 particles.  In almost all runs $N_{\rm
  rings}\!=\!1000$, however we created one simulation with $N_{\rm
  rings}\!=\!10,000$ (marked `HIRES' in the figure).  In all these
calculations, we find that the behavior of the disk particles was the
same.  Thus, we conclude that even in the highly damped
shear-dominated regime, our collisional code has adequate resolution
for all the cases we have studied.

The behavior of the embryos, on the other hand, only converged for
$N_{\rm tracer}\!\gtrsim\!10,000$.  \Red{There are two reasons why the
  number of tracers can effect the dynamical state of the embryos.
  First, the code could be} struggling to calculate accurately the
dynamical \Red{friction} of the tracers on the embryos.
Equation~\ref{eq_df} shows that the strength of dynamical friction
should not depend on the size of the tracer particle if $m_{\rm
  em}/m_{\rm tr}\gg 1$, but the question is how big does this ratio
have to be.  \Red{Second, the embryos could be artificially excited
  due to the viscous stirring from the large tracers.  In order for
  the code to behave correctly, the ratio of the tracer viscous
  stirring timescale to the viscous stirring timescale due to the
  embryos themselves must be greater than 1.  Combining G04's
  Equations 31 and 44, we expect that this ratio is proportional to
  $m_{\rm em}/m_{\rm tr}$, although again, it is not clear what this
  value needs to be in order to satisfy the timescale ratio
  constriant.  The fact that the code converges for $N_{\rm
    tracer}\!\gtrsim\!10,000$ implies that both the dynamical friction
  calculation is correct and the tracer viscous stirring is
  unimportant when $m_{\rm em}/m_{\rm tr} \Red{>} 150$.}  We abide by
this restriction in all the simulations that follow.

For the shear-dominated regime, G04 predicts (their Equation~77)
\begin{equation}
u \sim \frac{v_{\rm esc}}{\alpha^{3/2}} {\Sigma\over \sigma} {s\over R},
\end{equation}
where $\alpha$ is the angular size of the Sun as seen from the
embryo.  In addition, G04's Equation~110 says
\begin{equation}
v \sim v_{\rm esc}\sqrt{\alpha} {\Sigma\over \sigma}.
\end{equation}
Plugging in the appropriate values for this test, we find that G04
predicts that $u\!\sim\!2\times10^{-4}\,{\rm AU\, yr^{-1}}$ and
$v\!\sim\!7\times10^{-4}\,{\rm AU\, yr^{-1}}$.  In our simulation we
find $u\!\sim\!4\times10^{-4}\,{\rm AU\, yr^{-1}}$ and, after
convergence, $v\!\sim\!10^{-3}\,{\rm AU\, yr^{-1}}$, which is in very
good agreement with G04's analytic theory.

 \medskip\medskip
 
 In conclusion, in this subsection we presented a series of tests of
 SyMBA\_COL.  In all cases, the code reproduces the desired behavior.
 In addition, in those cases where direct comparison with G04's
 derivations is appropriate, there is reasonable quantitative
 agreement, within a factor of a few.  This level of agreement is
 about what one should expect given the order-of-magnitude nature of
 G04's development. Thus, in the remainder of the paper we employ the
 code to test C06's scenario for the early sculpting of the Kuiper
 belt.

\medskip
\medskip
\section{Systems with Five Ice Giants}
\label{sec_5-16}

In this section we use SyMBA\_COL to perform full dynamical
calculations of the scenario reviewed in \S{\ref{sec_chiang}}.  In
particular, for reasons discussed in \S{\ref{sec_chiang}}, we
concentrate on the phase when the ice giant system becomes unstable.
Thus, we start our systems with a {series of ice giants}, each of
which is $ {16}\mearth$, spread from 20 to $35\au$, which corresponds
to a spacing of roughly 5 Hills sphere.  In all cases, the initial
eccentricities and inclinations of the {ice giants were} very small
($\sim\!10^{-4}$), and Jupiter and Saturn were included on their
current orbits.

At the time of the instability, C06 predict that the mass of the
planetesimal disk should be about the same as the total mass of the
ice giants, i$.$e$.$ $ {80}\mearth$.  However, given the nature of
G04's derivations, this number is very uncertain.  Thus, we study a
range of disk masses: $m_{d}\!=\!40$, 80, 120, and $160\mearth$.  In
addition, it is uncertain what $s$ should be, {and thus we study two
  extremes: } $s=1\,$m and $1\,$cm.  The simulations {initially}
contain 2000 tracer particles which are spread from $16$ to $45\,$AU.
The initial eccentricities of the disk particles were set to
$\sim\!10^{-4}$, while the inclinations set to half of this value.
These parameters put our code in a regime where its validity and
convergence has been demonstrated in the tests in \S{\ref{sec_tests}}.

Except where noted, we integrated the systems for $10^8\,$yrs, with a
timestep of $0.4\,$yr.  As described in Levison \& Duncan~(2000),
SyMBA, and thus SyMBA\_COL, has difficulty handling close encounters
with the Sun.  Therefore, we remove from the simulation any object the
reaches a heliocentric distance less than $2\,$AU.  In all, we
performed 10 simulations.

\begin{figure}[h!]
\vglue 4.3truein
\includegraphics{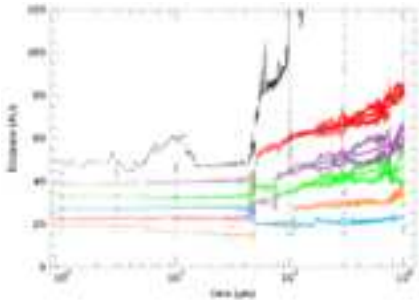}
\caption{\footnotesize \label{fig_aqQ1}
  {The temporal evolution of the \Red{five} ice giant simulation with
    $s=1\,$m and $m_{d}\!=\!80\mearth$.  The ice giants had an
    original mass of $16\,\mearth$.  We plot three curves for each
    planet, which is represented by a different color.  These curves
    show the semi-major axis, perihelion, and aphelion distances. The
    solid black curve illustrates the semi-major axis of the outermost
    disk particle.  The columns of black points show the semi-major
    axes of all the disk particles at various times.}}
\end{figure}

We start the discussion of our results with the $s=1\,$m runs.  In
particular, perhaps the best way to begin is with a detailed
description of the behavior on one particular simulation.  For this
purpose we chose the $m_{d}\!=\!80\mearth$ run since it is closest to
G04's nominal $\Sigma\!\sim\!\sigma$ situation.
Figure~{\ref{fig_aqQ1}} shows the temporal evolution of the nominal
simulation. In the figure each ice giant is represented by three
curves of the same color.  The curves show its semi-major axis,
perihelion, and aphelion distances.  The black curve shows the
semi-major axis of the outermost disk particle.  In addition to the
curves, there are columns of points.  These points show the semi-major
axes of all the disk particles at that particular time.

The system remains relatively quiescent for the first $500,000\,$yr.
During this time, the disk particles rearrange themselves so that a
large fraction of them are in the Trojan points of the ice giants.
This behavior can be seen in the two leftmost column of dots in the
figure.  In addition, two rings of particles form immediately interior
to and exterior to the embryos.  Originally the embryo growth rate is
large, the \Red{embryo mass increases $1.1\,\mearth$, on average,} in
the first $60,000\,$yr.  But after that time, very little growth
happens.  Again, this is due to the fact that most of the mass of the
disk is found in the Trojan points of the embryo and in isolated rings
where they are protected from the embryos.  At $550,000$ years the
system becomes mildly unstable and undergoes a `spreading event' that
moves the inner ice giant inward and the outer one outward.  Such
events are common in our simulations.  After this event the
eccentricities of the ice giants decrease presumably due to the fact
that they are further from one another.  For the next $\sim\!4\,$Myr
the system is stable.
  
During this period of relative quiescence the disk particles
concentrate in three main areas.  Roughly 30\% of the disk particles
can be found in a ring between the orange and red embryos.  This ring
therefore contains $22\,\mearth$!  It is very narrow as well --- only
$0.05\,$AU in width. Roughly the same amount of material can be found
in a ring at $47\,$AU, beyond all of the embryos.  This ring is also
very narrow with a width of only $\sim\!0.1\,$AU.  Finally, 29\% of
the disk can be found in the Trojan points of the green embryo.  This
implies that there is more mass in the Lagrange points than in the
embryo, itself.  The characteristics of these structures leaves us
wondering how physically realistic they are.  After all, our code
ignores fragmentation, which may be important as the ring forms.  In
addition, once the ring forms, the relative velocities are very small
and the surface densities are large, thus we might expect either
two-body accretion or a gravitational instability to form larger
objects.  We believe that the ring is an artifact of the simplistic
collisional physics that we inherited from G04 and they are probably
not physical.

At $4.3\,$Myr, the red and green embryos in the figure hit the 4:3
mean motion resonance with one another.  This destabilizes the embryos
and they undergo a series of scattering events with one another. A
large number of disk particles are released from their their storage
locations at this time.  This period of violence lasts for
$700,000\,$yrs, but eventually the dynamical friction caused by the
released disk particles is able to decouple the embryos from one
another.  Amusingly, the blue and orange embryos get trapped in the 1:1
mean motion resonance with one another, which lasts for over 5 million
years.

During the remainder of the simulation, the ice giants migrate
outward.  The disk particles that were liberated during the
instability are spread out enough that their collisional damping time
is longer than the time between encounters with the embryos.  Thus,
the planets hand the particles off to one another allowing a
redistribution of angular momentum.  This process, which is called
{\it planetesimal-driven migration} is well understood (see Levison et
al$.$~2006 for a review) since it was discovered over 20 years ago
(Fern\'andez \& Ip~1986) and has been studied as a possible
explanation of the resonant structure of the Kuiper belt
(Malhotra~1995, Hahn \& Malhotra~2005).

\begin{figure}[h!]
\vglue 4.2truein
\includegraphics{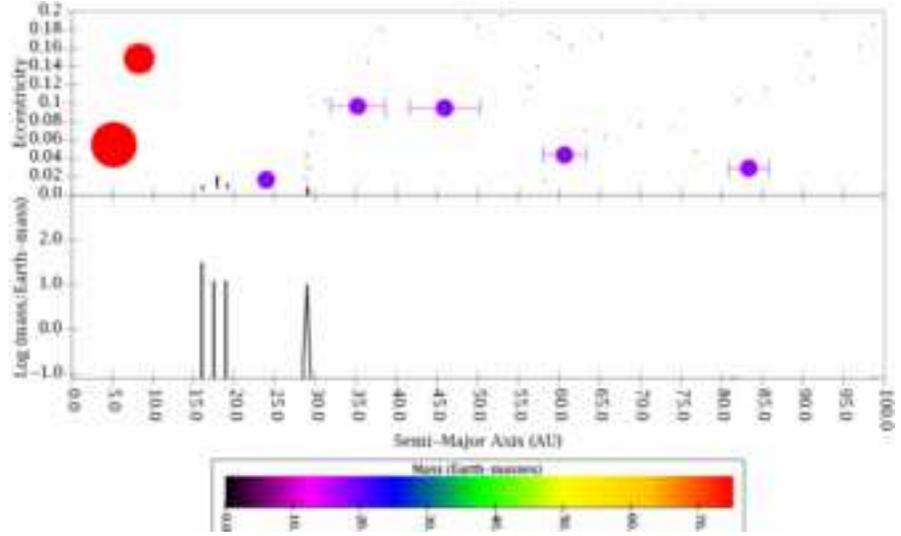}
\caption{\footnotesize \label{fig_final1m}
   {The final state of the 5 ice giant simulation with $s=1\,$m
    and $m_{d}\!=\!80\mearth$.  A) The eccentricity of an object as a
    function of its semi-major axes.  The disk particles are shown as
    black dots.  The planets are in color. the value of which is a
    function of the planet's mass.  The legend for the color is shown
    below the plot.  For the planets, the radius of the dot scales as
    the cube root of its mass.  The `error-bars' show the radial
    extent that the object travels as it orbits due to its
    eccentricity.  B) A histogram of the mass distribution of disk
    particles.  The width of the semi-major axis bins is $0.5\,$AU.
    Notice that almost all of the mass is concentrated in a few narrow
    rings.}}
\end{figure}

However, collisional damping still does play a role during this time.
As the planets migrate, four relatively high mass rings start to form.
At the end of the simulation the most massive of these contain
$32\,\mearth$!  Indeed, the final system is shown in
Figure~\ref{fig_final1m}.  Figure~\ref{fig_final1m}A shows the
eccentricity of objects as a function of their semi-major axis.  The
planets are in color, with the `error-bars' showing the range of
heliocentric distances that they travel as they orbit.  We find that
the inclinations, which are not shown, are roughly what one would
expect --- i$.$e$.$ the $\sin{(i)}$'s are roughly half the
eccentricities, This is a general result that we see in all the runs.

The size of the symbol in the figure scales as the mass of the planet
to the $1/3$ power.  The disk particles are shown in black.
Figure~\ref{fig_final1m}B presents a histogram of the mass of the disk
particles as a function of semi-major axis.  There are three important
things to note about the final system: 1) None of the ice giants was
ejected from the system.  2) There is a planet in a nearly circular
orbit at a large heliocentric distance (although it is close enough
that if it actually existed it would have been discovered long ago).
Although this system may not be finished evolving, it is very unlikely
that this planet will be removed.  3) Almost all the disk particles
survive.  Of the $80\mearth$ of material in the disk, $72\mearth$ are
still present at $100\,$Myr.  Almost all of this mass is found in four
massive rings interior to $30\,$AU.

\begin{figure}[h!]
\vglue 4.9truein
\includegraphics{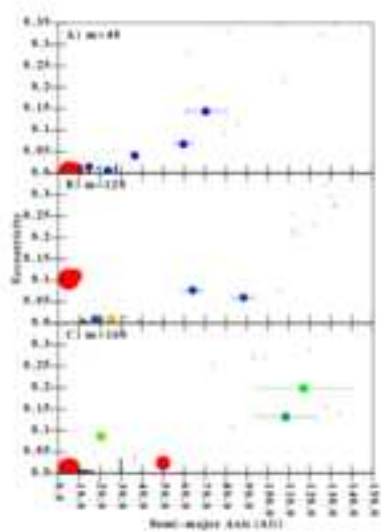}
\caption{\footnotesize \label{fig_finalallm}
  The final state of the three 5 ice giant simulations with $s=1\,$m
  not shown in Figure~\ref{fig_final1m}.  See
  Figure~\ref{fig_final1m}A for a description of this type of plot.
  A) $40\,\mearth$ disk.  B) $120\,\mearth$ disk.  C)
    $160\,\mearth$ disk.  }
\end{figure}

We performed 3 other simulations with $s\!=\!1\,$m and the stories for
these simulations are, for the most part, very similar.  The final
states of these systems are shown in Figure~\ref{fig_finalallm}.  This
figure is the same as Figure~\ref{fig_final1m}A, which fits in the
sequences between Figure~\ref{fig_finalallm}A and B.  The first thing
to note is that the entire planetary system went unstable in the
$m_d\!=\!160\,\mearth$ run.  In this case, a $75\,\mearth$ ring formed
in Saturn's 1:2 mean motion resonance that eventually drove up
Saturn's eccentricity until the Jupiter-Saturn system was disrupted.
We will ignore this run in the following analysis because it clearly
cannot represent what happened in the Solar System.  There were no
planetary ejections in any of these simulations in the remaining runs.
There was one merger in the $m_d\!=\!120\,\mearth$ run.  In all cases,
we find massive planets at large heliocentric distances, inconsistent
with the current Solar System and the expectations of G04 and C06.

As we described above, we are interested in setting up our initial
conditions so that they represent C06's hypothetical Solar System
immediately before the instability sets in, when the ice giants are
presumably almost fully formed.  That is reason we started with
systems where the initial mass of the ice giants was set to
$16\,\mearth$.  Thus, to test this assumption it is interesting to
look at the amount of mass accreted by the ice giants during our
calculations.  The colors in Figure~\ref{fig_final1m}A and
Figure~\ref{fig_finalallm} indicate the mass of the planet.  We found
that between 10\% and 16\% of the original disk mass is accreted by
the ice giants in our $s\!=\!1\,$m runs.  The amount of embryo growth
increases monotonically with disk mass.  The ice giants in the
$m_d\!=\!40\,\mearth$ run only accreted a total of $6.5\,\mearth$ of
planetesimals, while they grew roughly by a total of $13\,\mearth$ in
the $m_d\!=\!120\,\mearth$ simulation run.  The planets in this run
are probably too large to be considered good Uranus and Neptune
analogs.  It is also interesting to note that the more massive the
disk, the more excited the final system of ice giants is.  We think
that this is due to the fact that the more massive disks produce
larger planets, which, in turn, produce stronger mutual perturbations,
and thus stability is achieved only with wider orbital separations.

\begin{figure}[h!]
\vglue 4.5truein
\includegraphics{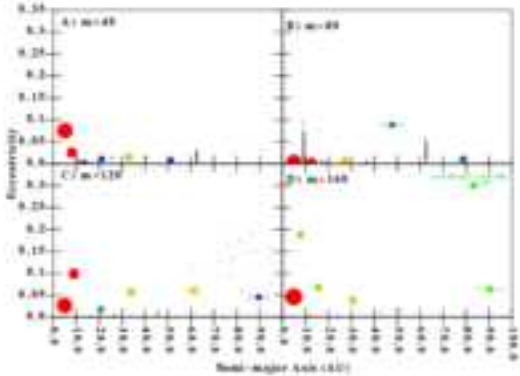}
\caption{\footnotesize \label{fig_finalallcm}
  The final state of the 5 ice giant simulations with $s=1\,$cm.
  These are the runs where the initial mass of the planets was
  $16\,\mearth$.  See Figure~\ref{fig_final1m}A for a description of
  this type of plot. A) $40\,\mearth$ disk.  B) $80\,\mearth$ disk.
  C) $120\,\mearth$ disk.  D) $160\,\mearth$ disk. The ice giant in
  the upper right of this panel should actually be at $a\!=\!161\,$AU
  and $e\!=\!0.48$.  It is plotted at its perihelion distance rather
  than semi-major axis.}
\end{figure}

We now turn our attention to the $s\!=\!1\,$cm runs --- the results of
which are shown in Figure~\ref{fig_finalallcm}.  The same basic
dynamics that we discussed above work with these systems and we get
the same basic results.  This includes the fact that Jupiter and
Saturn went unstable in the $m_d\!=\!160\,\mearth$ run.  The main
difference between the $m_d\!\leq\!120\,\mearth$ runs here and those
in the $s\!=\!1\,$m is that in these runs the ice giants accreted much
more of the disk material.  For example, the ice giants in the
$m_d\!=\!120\,\mearth$ run accreted a total of $37\,\mearth$ of
planetesimals in the $s\!=\!1\,$cm run, while they accreted
$13\,\mearth$ in the $s\!=\!1\,$m simulation.  This is probably due to
the fact that, at least at early times, collisions between the disk
particles are more frequent and thus the system remains cooler.  As a
result, the gravitational focusing factor of the ice giants remains
larger so that the accretion rate is higher. As a consequence, the
final ice giant systems that we obtain in these runs are, for the same
initial disk mass, more excited and spread out than those
constructed in the $s\!=\!1\,$m simulations
(Figure~\ref{fig_finalallm}).

In response to the large accretion rates in our $s\!=\!1\,$cm runs, we
decided to push our simulation back to an earlier time in order to
determine if we can produce ice giants on the order of the same size
as Uranus and Neptune for this value of $s$.  In particular, our
systems started with five embryos of $8.3\,\mearth$ each.  The orbits
of the embryos and the geometry of the disk was the same as in our
previous integrations.  We preformed 4 simulations with $m_d\!=43$,
85, 128, and $170\,\mearth$.  As with the previous cases, the orbits
of Jupiter and Saturn became unstable in the high disk mass.  \Red{Not
  surprisingly, we found a monotonic relationship between the initial
  disk mass and the amount of material the ice giants accreted from
  the disk in the remaining systems.  In particular, on average each
  ice giant accreted 2, 6, and $16\,\mearth$ in the runs with
  $m_d\!=\!43$, 85, and $128\,\mearth$, respectively.  Since we
  started with embryos of $8.3\,\mearth$, the $m_d\!=\!85\,\mearth$
  disk produces the best planets. However, we believe that larger disk
  masses would still probably acceptable if we were to increased $s$
  (thereby decreasing the damping).  On the other hand, since
  $s\!=\!1\,$cm is so extreme, we believe that we can rule out less
  massive disks (i$.$e$.$ $m_d\!\lesssim\!50\,\mearth$) --- disks this
  anemic are unlikely to form objects as massive as Uranus and
  Neptune.}

\begin{figure}[h!]
\vglue 4.5truein
\includegraphics{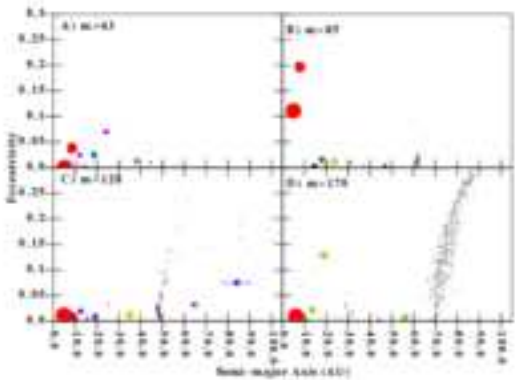}
\caption{\footnotesize \label{fig_finalallsm}
  The final state of the 5 ice giant simulations with $s=1\,$cm.
  These are the runs where the initial mass of the ice giants was
  $8.3\,\mearth$.  See Figure~\ref{fig_final1m}A for a description of
  this type of plot. A) $43\,\mearth$ disk.  B) $85\,\mearth$ disk.
  C) $128\,\mearth$ disk.  D) $170\,\mearth$ disk.}
\end{figure}

The final systems are shown in Figure~\ref{fig_finalallsm}.  \Red{As}
the planets grew, their orbits spread.  The final systems always had a
planet well beyond $30\,$AU.

\medskip
\medskip
\section{System with Earth-Mass Embryos}
\label{sec_6-sm}

Finally, in this section we briefly study the growth of the ice giants
from much smaller planetary embryos.  Our goal here is to make sure
that the initial conditions used above were not artificial in some
respect.  That is, given the order-of-magnitude nature of G04's
arguments, perhaps we start the above calculations in the wrong state.
For example, in all the above runs, we started with \Red{five} ice
giants, as suggested by C06.  However, perhaps the natural system
should contain \Red{four} such objects, rather than five, and then the
system might evolve as C06 suggested.  Thus, in this section we
present simulations where we start with small planetary embryos and
let the system evolve naturally.  Unfortunately, these simulations are
computationally expensive and thus we can only perform a couple of
test cases.

In particular, our systems started with 6 embryos of $1\,\mearth$
each, spread from 21 to $27\,$AU.  The initial eccentricities and
inclinations of these particles are very small ($\sim\!10^{-3}$).
These objects were embedded in a disk of 18,000 tracer particles
spread from 20 to $30\,$AU, with a total mass of $90\,\mearth$.  Two
simulations were done, one with $s\,=\,1\,$m and one with
$s\,=\,1\,$cm.

Performing a computation with 18,000 tracer particles is very CPU
intensive.  We needed such a large number of tracers to adequately
resolve the dynamical friction between the tracers and embryos, which,
as we explained above, requires that $m_{\rm em}/m_{\rm tr}\gg 150$.
However, as the embryos grow, such a large number of tracers were no
longer needed.  Thus, our plan was to continue the integrations until
the average embryo mass was $10\,\mearth$, after which we would remove
\Red{four} out of every \Red{five} tracer particles at random while
keeping the total mass of tracers constant.

\begin{figure}[h!]
\vglue 4.0truein
\includegraphics{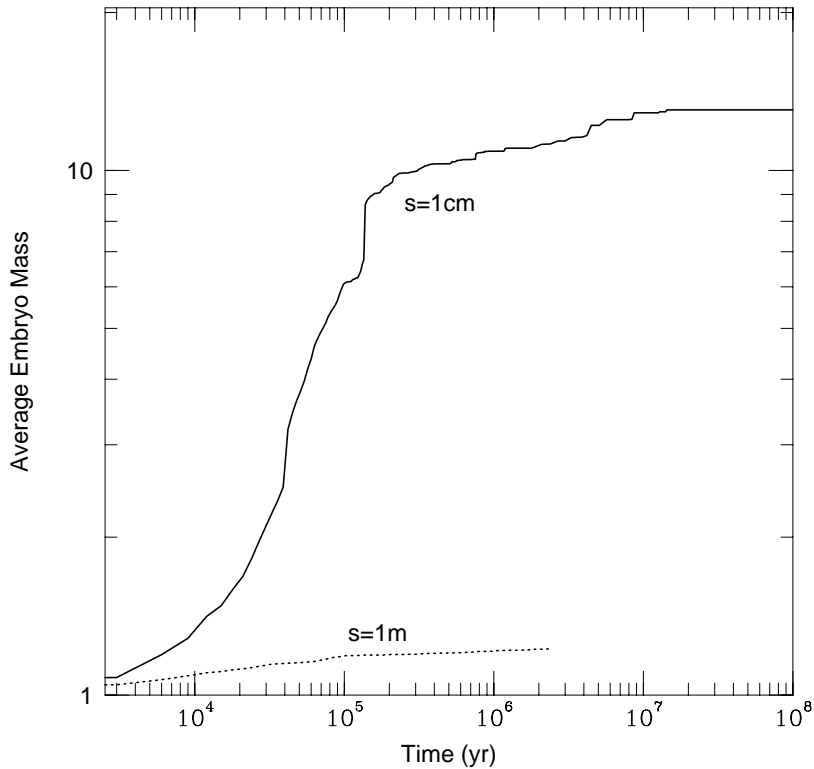}
\caption{\footnotesize \label{fig_s4_mass}
  The average mass of the embryos as a function of time in our
  simulations which start with 6 Earth-mass embryos.  The solid and
  dotted curves refer to the $s\!=\!1\,$cm and $s\!=\!1\,$m,
  respectively.  The $s\!=\!1\,$m runs was terminated at $2.5\,$Myr
  because of very small growth rates.}
\end{figure}

Figure~\ref{fig_s4_mass} shows the temporal evolution of the average
embryo mass in both the simulations.  The dotted curve is from the
$s\!=\!1\,$m.  Note that the growth rate is very small --- the average
mass of the embryos at $2.5\,$Myr was only $1.2\,\mearth$.  In
addition, during the last $500,000$ years of the simulation, the
growth rate was only $3\times 10^{-9}\,\mearth\,{\rm yr}^{-1}$.  We
terminated the simulation at this point because it was clear that this
simulation was no going to produce Uranus- and Neptune-sized planets
fast enough.  This is true because Uranus and Neptune, their ice giant
status notwithstanding, do each have several $M_\oplus$ of H and He in
their atmospheres. The most natural way to account for this is if
these planets finished their accretion in $\sim\!10^7$ years, before
the gas nebula was completely depleted (Haisch, Lada \& Lada 2001).

\begin{figure}[h!]
\vglue 4.0truein 
\includegraphics{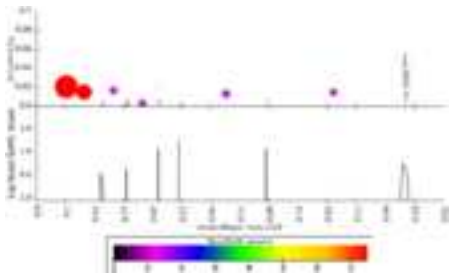}
\caption{\footnotesize \label{fig_s4_m_final}
  The final state of our $s\,=\,1\,$cm simulation with Earth-mass
  embryos. See Figure~\ref{fig_final1m} for a description of this type
  of plot.}
\end{figure}

As Figure~\ref{fig_s4_mass} shows, however, the $s\!=\!1\,$cm runs,
indeed, produce reasonable Uranus and Neptune analogs within $10\,$Myr.
In fact, by this time, the embryos have an average mass of
$12.4\,\mearth$.  The final system is shown in
Figure~\ref{fig_s4_m_final}.  At $100\,$Myr, this run has four ice
giants ranging in mass from 10.6 to $14.9\,\mearth$.  Recall that the
system started with six embryos.  There were two mergers that reduced
the number to four.  Thus, none of the ice giants were ejected.  Two
of the resulting planets have semi-major axes beyond the current orbit
of Neptune.  Of particular note is the $10.6\,\mearth$ ice giant on an
orbit with a semi-major axis of $51\,$AU and an eccentricity of 0.01.
This system suffers from another problem as well.  Less than 50\% of
the tracers were accreted by the planets.  Thus, there is still
$46\,\mearth$ of material concentrated in narrow rings through the
outer planetary system.

\medskip
\medskip
\section{Conclusions}
\label{sec_conc}

C06 has recently proposed a new and innovative scenario for the
primordial sculpting of the Kuiper belt.  The idea is based on a
recent pair of papers, G04a and G04b, that, based on
order-of-magnitude analytic arguments, predicted that originally
\Red{$\sim\!5$} planets began to grow between $\sim\!20$ and
$\sim\!40\,$AU.  As these planets grew to masses of
$\sim\!15\,M_\oplus$ their orbits went unstable, \Red{some} of them
were ejected, leaving Uranus and Neptune in their current orbits.  C06
argued that this violent process could explain the \Red{the currently
  observed dynamical excitation of the Kuiper belt}.  Like G04, the
C06 scenario was not tested with numerical simulations, but was solely
supported by back-of-the-envelope analytic estimates.

Therefore, here we presented a series of integrations intended to
simulate numerically C06's scenario.  We performed 12 simulations
starting at the stage where the 5 ice giants are predicted to become
unstable.  In these simulations we varied the mass contained in the
background disk, the size of the disk particles (either $1\,$m or
$1\,$cm), and the initial mass of the ice giants.  We found that in
the simulations in which the mass of the disk $\gtrsim\!160\,\mearth$,
the orbits of Jupiter and Saturn were not stable.  \Red{In addition,
  we can rule out systems with disk masses $\lesssim\!50\,\mearth$
  because our simulations that start with $8.3\,\mearth$ ice giants
  show that these disks are unlikely to produce planets as large as
  Uranus and Neptune.}

\Red{In the runs where Jupiter and Saturn were stable,} contrary to
G04's conjecture, planetary ejection \Red{almost} never occurs.
Instead, we found that the planetary systems spread and thus all our
final systems contained a planet in an orbit far beyond the current
orbit of Neptune (but still at a distance at which it would not escape
detection).  All our systems had at least one planet beyond $50\,$AU.
Indeed, the semi-major axis of the outermost giant planet in these
systems ranged from 52 to $90\,$AU.

\Red{Obviously, we have not been able to model all possible cases
  since parameter space is large and these calculations are expensive.
  However, we think that it is possible to speculate how these results
  would change if we expanded our coverage of parameter space.  It
  would help if we were to reduce the number of ice giants to four, or
  even three, because we would need to eject fewer planets.  However,
  since we found only one ejection in our entire set of simulations
  (only including those runs where Jupiter and Saturn were stable), we
  do not believe that this would solve the problem.}

\Red{It might also help if we were to make the initial planetary
  system more compact and thus make it more likely that the ice giants
  would evolve onto Saturn-crossing orbits.  However, we find that for
  the disk masses we have studied, Saturn only has a 17\% chance of
  ejecting an ice giant that is crossing its orbit.  Therefore, a more
  compact system probably will not solve the problem. This conclusion
  is at odds with the results of Thommes et al$.$~(1999; 2002), where
  ice giants originally in compact planetary systems were commonly
  ejected.  We believe that this difference is due to the fact that
  our disks are collisionally active.  In the Thommes et al$.$
  simulations, disk particles are very quickly removed as the result
  of gravitational interaction with the planets.  This does not occur
  in our simulations because collisions keep the disk in place.  This
  discussion leads, however, to the final way in which we could
  increase the number of ejections --- we could increase $s$ thereby
  making these simulations more like those in Thommes et al$.$
  However, we find that even in our $s\!=\!1\,$m runs, the accretion
  rate is so small that in a situation where $s$ is large, Uranus and
  Neptune would probably not grow (see Levison \& Stewart~2001), at
  least by G04's mechanism.}

\Red{As a case in point}, we performed two simulations that initially
contained six Earth-mass embryos \Red{embedded in a $90\,\mearth$
  disk}.  One run had $s\,=\,1\,$m, and the other had $s\,=\,1\,$cm.
In the $s\,=\,1\,$m run, the growth rates were too small to allow
Uranus-Neptune analogs to form in a reasonable amount of time.
However, four ice-giants with masses between 10.6 to $14.9\,\mearth$
formed in the $s\,=\,1\,$cm run.  The outermost of these had a
semi-major axis of $51\,$AU and an eccentricity of 0.01.

All the simulations we have performed show the same basic behavior ---
the system spreads during the growth and dynamical evolution of the
ice giants.  Planetary ejections are rare.  \Red{For reasons described
  above,} we believe these results are generic enough to be universal.
Therefore, we believe that it is safe to rule out the C06's scenario
for the sculpting of the Kuiper belt, as well as G04's scenario for
the formation of Uranus and Neptune.

We think that the problem with G04 is not in the derivation of the
various estimates, but rather in some of simplifying assumptions that
they were forced to employ to make the problem analytically tractable.
Indeed, on microscopic, short-term, scales, we were able to reproduce
much of the behavior that G04 predicted (see \S{\ref{sec_code}}).  In
the case of the problem we address in this paper, G04's assumption
that the surface density of the disk particles remains smooth and
uniform is probably at fault, since we find that the formation of
rings and gaps actually dominates the dynamics.  Having said this, we
must remind the reader that we adopted many of G04's simplifying
assumption ourselves, and if this mechanism for planet formation is to
be further explored, these should be more fully tested.

\medskip\medskip

\acknowledgments HFL is grateful for funding from NASA's Origins and
PGG programs.  AM acknowledges funding from the french National
Programme of Planetaology (PNP).  We would also like to thank John
Chambers, Eiichiro Kokubo, and an anonymous third person for acting as
referees on this manuscript.  Their comments were much appreciated.
Finally, we thank Glen Stewart for useful discussions.

\section*{References}

\begin{itemize}
\setlength{\itemindent}{-30pt}
\setlength{\labelwidth}{0pt}

\item[] Allen, R.~L., Bernstein, G.~M., Malhotra, R.\ 2001.\ The Edge
  of the Solar System.\ Astrophysical Journal 549, L241-L244.

\item[] Allen, R.~L., Bernstein, G.~M., Malhotra, R.\ 2002.\ 
  Observational Limits on a Distant Cold Kuiper Belt.\ Astronomical
  Journal 124, 2949-2954.
  
\item[] Bernstein, G.~M., Trilling, D.~E., Allen, R.~L., Brown, M.~E.,
  Holman, M., Malhotra, R.\ 2004.\ The Size Distribution of
  Trans-Neptunian Bodies.\ Astronomical Journal 128, 1364-1390.
  
\item[] Binney, J., Tremaine, S.\ 1987.\ Galactic dynamics.\ 
  Princeton, NJ, Princeton University Press, 1987.
  
\item[] Brouwer, D., Clemence, G.~M.\ 1961.\ Methods of celestial
  mechanics.\ New York: Academic Press, 1961 .

\item[] Brown, M.~E.\ 2001.\ The Inclination Distribution of the
  Kuiper Belt.\ Astronomical Journal 121, 2804-2814.
  
\item[] Chambers, J.\ 2006.\ A semi-analytic model for oligarchic
  growth.\ Icarus 180, 496-513.
  
\item[] Chandrasekhar, S.\ 1943.\ Dynamical Friction. I. General
  Considerations: the Coefficient of Dynamical Friction..\ 
  Astrophysical Journal 97, 255.
  
\item[] Chiang, E.~I., Brown, M.~E.\ 1999.\ Keck Pencil-Beam Survey
  for Faint Kuiper Belt Objects.\ Astronomical Journal 118, 1411-1422.
  
\item[] Chiang, E., Lithwick, Y., Murray-Clay, R., Buie, M., Grundy,
  W., Holman, M. 2006. A Brief History of Trans-Neptunian Space.\ 
  Protostars and Planets V.  In Press.
  
\item[] Duncan, M.~J., Levison, H.~F., Lee, M.~H.\ 1998.\ A Multiple
  Time Step Symplectic Algorithm for Integrating Close Encounters.\ 
  Astronomical Journal 116, 2067-2077.
  
\item[] Fernandez, J.~A., Ip, W.-H.\ 1984.\ Some dynamical aspects of
  the accretion of Uranus and Neptune --- The exchange of orbital
  angular momentum with planetesimals.\ Icarus 58, 109-120.
  
\item[] Gladman, B., Kavelaars, J.~J., Petit, J.-M., Morbidelli, A.,
  Holman, M.~J., Loredo, T.\ 2001.\ The Structure of the Kuiper Belt:
  Size Distribution and Radial Extent.\ Astronomical Journal 122,
  1051-1066.
  
\item[] Goldreich, P., Lithwick, Y., Sari, R.\ 2004.\ Final Stages of
  Planet Formation.\ Astrophysical Journal 614, 497-507.

\item[] Goldreich, P., Lithwick, Y., Sari, R.\ 2004.\ Planet Formation
  by Coagulation: A Focus on Uranus and Neptune.\ Annual Review of
  Astronomy and Astrophysics 42, 549-601.
  
\item[] Gomes, R.~S.\ 2003.\ The origin of the Kuiper Belt
  high-inclination population.\ Icarus 161, 404-418.
  
\item[] Gomes, R., Levison, H.~F., Tsiganis, K., Morbidelli, A.\ 
  2005.\ Origin of the cataclysmic Late Heavy Bombardment period of
  the terrestrial planets.\ Nature 435, 466-469.

\item[] Hahn, J.~M., Malhotra, R.\ 1999.\ Orbital Evolution of Planets
  Embedded in a Planetesimal Disk.\ Astronomical Journal 117,
  3041-3053.

\item[] Hahn, J.~M., Malhotra, R.\ 2005.\ Neptune's Migration into a
  Stirred-Up Kuiper Belt: A Detailed Comparison of Simulations to
  Observations.\ Astronomical Journal 130, 2392-2414.
  
\item[] Haisch, K.~E., Jr., Lada, E.~A., Lada, C.~J.\ 2001.\ Disk
  Frequencies and Lifetimes in Young Clusters.\ Astrophysical Journal
  553, L153-L156.

\item[] Jewitt, D., Luu, J., Chen, J.\ 1996.\ The Mauna
  Kea-Cerro-Tololo (MKCT) Kuiper Belt and Centaur Survey.\ 
  Astronomical Journal 112, 1225.
  
\item[] Kenyon, S.~J., Luu, J.~X.\ 1998.\ Accretion in the Early
  Kuiper Belt. I. Coagulation and Velocity Evolution.\ Astronomical
  Journal 115, 2136-2160.
  
\item[] Kenyon, S.~J., Luu, J.~X.\ 1999.\ Accretion in the Early Outer
  Solar System.\ Astrophysical Journal 526, 465-470.
  
\item[] Kokubo, E., Ida, S.\ 1996.\ On Runaway Growth of
  Planetesimals.\ Icarus 123, 180-191.
  
\item[] Kokubo, E., Ida, S.\ 1998.\ Oligarchic Growth of
  Protoplanets.\ Icarus 131, 171-178.
  
\item[] Levison, H.~F., Lissauer, J.~J., Duncan, M.~J.\ 1998.\ 
  Modeling the Diversity of Outer Planetary Systems.\ Astronomical
  Journal 116, 1998-2014.
  
\item[] Levison, H.~F., Duncan, M.~J.\ 2000.\ Symplectically
  Integrating Close Encounters with the Sun.\ Astronomical Journal
  120, 2117-2123.
  
\item[] Levison, H.~F., Stern, S.~A.\ 2001.\ On the Size Dependence of
  the Inclination Distribution of the Main Kuiper Belt.\ Astronomical
  Journal 121, 1730-1735.
  
\item[] \Red{Levison, H.~F., Stewart, G.~R.\ 2001.\ Remarks on
    Modeling the Formation of Uranus and Neptune.\ Icarus 153,
    224-228.}
  
\item[] Levison, H.~F., Morbidelli, A.\ 2003.\ The formation of the
  Kuiper belt by the outward transport of bodies during Neptune's
  migration.\ Nature 426, 419-421.
  
\item[] Levison, H.~F., Morbidelli, A., Gomes, R., \& Backman, D.
  2006. Planet Migration in Planetesimal Disks.\ Protostars and
  Planets V.  In Press.
  
\item[] Malhotra, R.\ 1995.\ The Origin of Pluto's Orbit: Implications
  for the Solar System Beyond Neptune.\ Astronomical Journal 110, 420.
  
\item[] Miller, R.~H.\ 1978.\ Numerical experiments on the stability
  of disklike galaxies.\ Astrophysical Journal 223, 811-823.
  
\item[] Morbidelli, A., Brown, M.~E., Levison, H.~F.\ 2003.\ The
  Kuiper Belt and its Primordial Sculpting.\ Earth Moon and Planets
  92, 1-27.
  
\item[] Murray, C.~D., Dermott, S.~F.\ 2000.\ Solar System Dynamics.\ 
  Solar System Dynamics, by C.D.~Murray and S.F.~Dermott.~ ISBN
  0521575974.

\item[] Morbidelli, A., Jacob, C., Petit, J.-M.\ 2002.\ Planetary
  Embryos Never Formed in the Kuiper Belt.\ Icarus 157, 241-248.

\item[] Morbidelli, A.\ 2005.\ Origin and Dynamical Evolution of
  Comets and their Reservoirs.\ ArXiv Astrophysics e-prints
  arXiv:astro-ph/0512256.
  
\item[] Ohtsuki, K., Ida, S.\ 1990.\ Runaway planetary growth with
  collision rate in the solar gravitational field.\ Icarus 85,
  499-511.
  
\item[] Rafikov, R.~R.\ 2004.\ Fast Accretion of Small Planetesimals
  by Protoplanetary Cores.\ Astronomical Journal 128, 1348-1363.
  
\item[] Skeel, R.~D., \& Biesiadecki, J.~J.~1994. Ann.\ Numer.\ Math.,
  1, 191.
  
\item[] Stern, S.~A.\ 1996.\ On the Collisional Environment, Accretion
  Time Scales, and Architecture of the Massive, Primordial Kuiper
  Belt..\ Astronomical Journal 112, 1203.
  
\item[] Stern, S.~A., Colwell, J.~E.\ 1997.\ Accretion in the
  Edgeworth-Kuiper Belt: Forming 100-1000 KM Radius Bodies at 30 AU
  and Beyond..\ Astronomical Journal 114, 841.
  
\item[] Tegler, S.~C., Romanishin, W.\ 2003.\ Resolution of the kuiper
  belt object color controversy: two distinct color populations.\ 
  Icarus 161, 181-191.

\item[] \Red{ Thommes, E.~W., Duncan, M.~J., Levison, H.~F.\ 1999.\ 
    The formation of Uranus and Neptune in the Jupiter-Saturn region
    of the Solar System.\ Nature 402, 635-638. }
  
\item[] \Red{Thommes, E.~W., Duncan, M.~J., Levison, H.~F.\ 2002.\ The
    Formation of Uranus and Neptune among Jupiter and Saturn.\ 
    Astronomical Journal 123, 2862-2883.}

\item[] Thommes, E.~W., Duncan, M.~J., Levison, H.~F.\ 2003.\ 
  Oligarchic growth of giant planets.\ Icarus 161, 431-455.

\item[] Trujillo, C.~A., Brown, M.~E.\ 2001.\ The Radial Distribution
  of the Kuiper Belt.\ Astrophysical Journal 554, L95-L98.
  
\item[] Trujillo, C.~A., Jewitt, D.~C., Luu, J.~X.\ 2001.\ Properties
  of the Trans-Neptunian Belt: Statistics from the
  Canada-France-Hawaii Telescope Survey.\ Astronomical Journal 122,
  457-473.
  
\item[] Trujillo, C.~A., Brown, M.~E.\ 2002.\ A Correlation between
  Inclination and Color in the Classical Kuiper Belt.\ Astrophysical
  Journal 566, L125-L128.
  
\item[] Tsiganis, K., Gomes, R., Morbidelli, A., Levison, H.~F.\ 
  2005.\ Origin of the orbital architecture of the giant planets of
  the Solar System.\ Nature 435, 459-461.
  
\item[] Wisdom, J., Holman, M.\ 1991.\ Symplectic maps for the n-body
  problem.\ Astronomical Journal 102, 1528-1538.

\item[] Youdin, A.~N., Shu, F.~H.\ 2002.\ Planetesimal Formation by
  Gravitational Instability.\ Astrophysical Journal 580, 494-505.

\item[]  Youdin, A.~N., 
Chiang, E.~I.\ 2004.\ Particle Pileups and Planetesimal Formation.\ 
Astrophysical Journal 601, 1109-1119.

\end{itemize}

\clearpage

\end{document}